# Mass of the Mars-sized Exoplanet Kepler-138 b from Transit Timing

Daniel Jontof-Hutter[1,2], Jason F. Rowe[2,3], Jack J. Lissauer[2], Daniel C. Fabrycky[4,5], and Eric B. Ford[1]

**Extrasolar planets that pass in front of their host star (transit) cause a temporary decrease in the apparent brightness of the star once per orbit, providing a direct measure of the planet's size and orbital period. In some systems with multiple transiting planets, the times of the transits are measurably affected by the gravitational interactions between neighbouring planets[1,2]. In favorable cases, the departures from Keplerian orbits implied by the observed transit times permit planetary masses to be measured, which is key to determining bulk densities[3]. Characterizing rocky planets is particularly difficult, since they are generally smaller and less massive than gaseous planets. Thus, few exoplanets near Earth's size have had their masses measured. Here we report the sizes and masses of three planets orbiting Kepler-138, a star much fainter and cooler than the Sun. We measure the mass of the Mars-sized inner planet based on on the transit times of its neighbour and thereby provide the first density measurement for an exoplanet smaller than Earth. The middle and outer planets are both slightly larger than Earth. The middle planet's density is similar to that of Earth, while the outer planet is less than half as dense, implying that it contains a greater portion of low density components such as $H_2O$ and/or $H_2$.**

NASA's *Kepler* mission has discovered thousands of candidate transiting exoplanets, with a wide range of planetary sizes[4,5,6]. A small fraction of these planets have had their masses characterized, by either radial velocity spectroscopy (RV) or via transit timing. The latter probes the gravitational perturbations between planets in multi-planet systems by precisely measuring transit times and fitting dynamical models to the observed transit timing variations (TTV)[1,2]. Both RV and TTV signals are larger for more massive planets, improve with greater planetary masses, although the two techniques sample different populations of exoplanets. The RV technique measures the motion of a host star induced by its planet's gravity, and hence the signal declines with increasing orbital distance. The majority of planets with mass determinations via RV from *Kepler*'s dataset have orbital periods shorter than one week. For Kepler-discovered planets characterized as rocky by this method, the orbital periods are all less than one day. Of the RV detections, Kepler-78 b has the lowest mass, (1.7 $M_{Earth}$), and the shortest orbital period- 0.35 days[7,8]. Characterizing planets by transit timing is quite complementary to RV because transit timing is very sensitive to perturbations between planets that are closely spaced or near orbital resonances[3,9]. Note that most systems with detected TTVs are not in resonance, but rather near enough to resonance for the perturbations to be coherent for many orbital periods, while also far enough from resonance that the planetary conjunctions cycle around the orbit plane within the four-year *Kepler* baseline. For near-resonant pairs, the TTVs of neighbouring planets frequently take the form of anti-correlated sinusoids over many orbits[10,11,12,13,14] or the sum of sinusoids where a planet is perturbed by two neighbours[15]. The majority of TTV detections have been found near first-order mean-motion resonances, such that planet pairs have an orbital period ratio near *j:j*-1 where *j* is an integer. Near first-order resonances generally cause stronger TTV signals than second-order resonances, although much depends on how close the planet pairs are to resonance and how eccentric (non-circular) their orbits are, as well as the planets' masses. Eccentricity causes the orbital speed to vary during the orbit, and the distance between the planets at conjunction to vary with the position of the conjunction.

[1] Department of Astronomy, Pennsylvania State University, Davey Lab, University Park, PA 16802, USA;
[2] NASA Ames Research Center, Moffett Field, CA 94035, USA;
[3] SETI Institute, 189 N Bernardo Ave, Mountain View, CA 94043, USA;
[4] Department of Astronomy and Astrophysics, University of Chicago, 5640 South Ellis Avenue, Chicago, IL 60637, USA
[5] Alfred P. Sloan Fellow

The bulk of planets with mass characterizations from the *Kepler* sample using TTVs so far have orbital periods ranging from ~10-100 days. These are generally low density[15,16,17,18] and likely possess deep atmospheres, with the exception of the rocky planet Kepler-36 b[19].

Kepler-138 (formerly known as KOI-314) hosts three validated transiting planets[20]. The orbital periods of Kepler-138's three planets are given in Table 1. Kepler-138 c and d orbit near a second order mean motion resonance (5:3), whilst 'b' and 'c' orbit near the 4:3 first order resonance.

Using transit times up to the fourteenth quarter of the *Kepler* mission, two of the three known planets orbiting Kepler-138[20], have been confirmed and characterized with TTVs[21]. The derived parameters in that work suggested that the outer planet, Kepler-138 d, has a density so low that it must have a significant hydrogen/helium gaseous envelope[21].

Using the complete *Kepler* dataset for Kepler-138, we have detected TTVs for all three planets. We describe our procedure to measure transit times in Methods, and list our transit time measurements in Extended Data Table 1. TTVs are expressed as the difference between the observed transit times and the calculated linear fit to the transit times.

Modeling Kepler-138 as a three-planet system, we measure the masses of all three planets, the super-Earth sized Kepler-138 c and d, as well as Kepler-138 b, which at 0.52 $R_{Earth}$ is roughly the size of Mars.

We performed dynamical fits by calculating the orbits of the three planets around the star, modeling the orbits as co-planar, since all three planets are transiting, Kepler's multi-planet systems are known to have small mutual inclinations, and we demonstrate that allowing mutual inclinations has little effect on our results for the planet masses (see Extended Data). Our model parameters for each planet are the orbital period *P*, the time $T_0$ of the first transit after our chosen epoch, the components of the eccentricity vector (*ecosω* and *esinω*, where *ω* is the angle between the sky-plane and the orbital pericenter of the planet), and $M_p/M_{star}$, the mass of the planet relative to the host star, which we express as $(M_p/M_{Earth})(M_{Sun}/M_{star})$ throughout. We perform Bayesian parameter estimation for these 15 model parameters using Differential Evolution Markov Chain Monte-Carlo.

We report the properties of the star and planets in Table 1, and details of the parameter estimation algorithm, priors and statistical models in Methods.

We measure the mass of Kepler-138 b to be $0.066^{+0.059}_{-0.037} M_{Earth}$, where uncertainties denote the 68.3% confidence limits. The 95.4% interval spans $0.011-0.170\ M_{Earth}$. The robustness of this result against outlying transit times and mutual inclinations is demonstrated in Methods.

The posterior probability for the inner planet having non-zero mass is between 99.82% and 99.91% (depending on the choice of prior for eccentricity), i.e., equivalent to a "3-sigma" detection. This calculation is based on the Savage-Dickey density ratio for calculating the Bayes factor, which fully accounts for posterior width and shape, including asymmetries and non-Gaussianity as described in Methods.

Kepler-138 b is by far the smallest exoplanet, both by radius and mass, to have a density measurement. Thus it opens a new regime to physical study. It is likely to become the prototype for a class of small close-in planets that could be common. The prospect of further constraints on this planet are excellent: NASA's Transiting Exoplanet Survey Satellite should be able to measure transit times for the two largest planets, improving constraints on the dynamical model for all 3 planets. ESA's *Plato* mission will continue this process and ground-based measurements are also possible. These future observations, plus more accurate stellar classification using the distance to the star measured by the *Gaia* mission, will further improve the characterization of this system, especially the inner planet.

Our measurements of the mass and density of the small inner planet Kepler-138 b are consistent with various compositions and formation locations. If future observations imply that the planet is less dense than rock, then the only physically and cosmogonically plausible low-density constituents are $H_2O$ and/or other astrophysical ices, which could only have condensed far from the star. This would be the first definitive evidence for substantial inward orbital migration of a small planet.

For the two outer planets, Kepler-138 c and d, we find a lower mass ratio between these two planets than previous work[21]. This is not surprising, since planet 'b's perturbations explain part of the TTVs observed in planet 'c', which were previously attributed solely to perturbations by planet 'd'. Nevertheless, the mass ratios between each of these planets and their host star remains consistent with published results[21]. We find higher densities for both of these planets than previous work, due to our improved stellar properties, particularly the higher stellar density and consequent smaller stellar radius.

Previous estimates of the size and mass of the outer planet Kepler-138 d implied that the planet possessed a hydrogen-rich atmosphere[21], which is difficult to explain with our current understanding of the accretion and retention of light gases from low-mass planets orbiting close to their star[28,29,30]. Our new measurements could be explained by a composition of rock and $H_2O$. A rock and $H_2O$ planet would be more stable against mass loss, and would imply that the planet formed at a greater distance from the star and migrated.

**Acknowledgements**

D.J. acknowledges support through the NASA Postdoctorial Program and funding from the Center for Exoplanets and Habitable Worlds. J.F.R. acknowledges NASA grant NNX14AB82G issued through the Kepler Participating Scientist Program. D.C.F. was supported by the NASA Kepler Participating Scientist Program award NNX14AB87G. E.B.F. was supported in part by NASA Kepler Participating Scientist Program award NNX14AN76G and NASA Exoplanet Research Program award NNX15AE21G. The Center for Exoplanets and Habitable Worlds is supported by the Pennsylvania State University, the Eberly College of Science, and the Pennsylvania Space Grant Consortium.

**Author Contributions**

**D. Jontof-Hutter** led the research effort to model the transit timing variations, constrain planetary masses, and wrote the manuscript. **J. Rowe** measured transit times from the *Kepler* dataset, characterized the host star using spectral follow-up of the target and constraints from the transits and edited the manuscript. **J. Lissauer** led the interpretation effort, assisted in the dynamical study and writing the manuscript. **D. Fabrycky** wrote software to simulate planetary transits, assisted in interpreting results and edited the manuscript. **E. Ford** assisted in the development of statistical methodologies and robustness tests for the TTV modeling and edited the manuscript.

**Author Information**
Reprints and permissions information is available at www.nature.com/reprints.
Correspondence and requests for materials should be addressed to D. Jontof-Hutter
dxj14@psu.edu.

| Stellar Parameters | Planet | Period (days) | $T_0$ (BJD-2,454,900) | $e\cos\omega$ | $e\sin\omega$ | $\dfrac{M_p}{M_{Earth}}\dfrac{M_{Sun}}{M_{star}}$ |
|---|---|---|---|---|---|---|
| $M_{star}$: 0.521 +/- 0.055 $M_{Sun}$ | b | $10.3126^{+0.0004}_{-0.0006}$ | $788.4142^{+0.0027}_{-0.0027}$ | $-0.011^{+0.096}_{-0.140}$ | $-0.024^{+0.075}_{-0.135}$ | $0.13^{+0.12}_{-0.08}$ |
| $R_{star}$: 0.442 +/- 0.024 $R_{Sun}$ | c | $13.7813^{+0.0001}_{-0.0001}$ | $786.1289^{+0.0005}_{-0.0005}$ | $-0.015^{+0.086}_{-0.126}$ | $-0.020^{+0.064}_{-0.117}$ | $3.85^{+3.77}_{-2.30}$ |
| $T_{eff}$: 3841 +/- 49 K | d | $23.0881^{+0.0009}_{-0.0008}$ | $796.6689^{+0.0013}_{-0.0013}$ | $-0.037^{+0.060}_{-0.092}$ | $-0.057^{+0.674}_{-0.387}$ | $1.28^{+1.36}_{-0.78}$ |
| Density: 9.5 +/- 2.2 g cm$^{-3}$ | Planet | $\dfrac{R_p}{R_{star}}$ | $M_p$ ($M_{Earth}$) | $R_p$ ($R_{Earth}$) | Density (g cm$^{-3}$) | Incident Flux (rel. to Earth) |
| [Fe/H]: -0.280 +/- 0.099 | b | 0.0108 +/- 0.0003 | $0.066^{+0.059}_{-0.037}$ | 0.522 +/-0.032 | $2.6^{+2.4}_{-1.5}$ | 6.81 +/- 0.84 |
| log(g) (cm s$^{-2}$): 4.886 +/0.055 | c | 0.0247 +/- 0.0005 | $1.970^{+1.912}_{-1.120}$ | 1.197 +/- 0.070 | $6.2^{+5.8}_{-3.4}$ | 4.63 +/- 0.57 |
|  | d | 0.0251 +/- 0.0007 | $0.640^{+0.674}_{-0.387}$ | 1.212 +/- 0.075 | $2.1^{+2.2}_{-1.2}$ | 2.32 +/- 0.29 |

**Table 1: Stellar and planetary parameters for the Kepler-138 system**. The left column lists our adopted stellar parameters. The upper right panel lists the solutions for the parameters we have explored with dynamical modeling. The lower right panel shows our adopted physical characteristics for the three planets.

Figure 1: Transit Timing Variations of the three planets orbiting Kepler-138. In black are the differences between measured transit times and a calculated linear fit to the transit times, with 1σ uncertainties shown as error-bars. Grey points mark the difference between the simulated transit times based on the best-fit dynamical model and a linear fit to the transit times. Panels 'a', 'b' and 'c' display the TTVs of Kepler-138 b, c and d respectively.

Figure 2: Mass-Radius diagram of well-characterized planets smaller than 2.1 $R_{Earth}$. Prior exoplanet characterizations are shown as grey points[6,7,9,16,19,22,23,24,25,26,27]. Black points from left to right are Mercury, Mars, Venus and Earth. Red data points are our results for Kepler-138. Open circles mark previously measured masses for Kepler-138 c and d[21]. Error bars mark published 1σ uncertainties for the planets of Kepler-138 and published masses and radii of all other characterized exoplanets within this size range. The curves mark bulk densities of 1, 3 and 10 g cm$^{-3}$.

**Methods**
We have used all available short cadence *Kepler* data and long cadence data wherever short cadence are unavailable to complete the dataset for 17 quarters. We list the transit times for each planet in Extended Data (ED) Table 1. Throughout, we express times since Barycentric Julian Day (BJD)- 2,454,900.

**Photometric Transit and Stellar Models**

From the light curve, we filtered instrumental and astrophysical effects that are independent of planetary transits. To each segment of the photometric time series, we fitted a cubic polynomial of width 2 days, centered on the time of each measurement[20]. We excluded measurements taken within 1 transit-duration (defined as the time from first to last contact) of the measured center of the transit and extrapolate the polynomial to estimate corrections during transits. This process strongly filters astrophysical signals with timescales of approximately 2 days, which could affect the shape of a planetary transit. We also excluded measurements for which the associated segment has gaps longer than 2.5 hours.

We fitted the detrended *Kepler* light curve using a transit model for quadratic limb-darkening[31] and non-circular Keplerian orbits. We stacked transits of each planet with corrections for the measured TTVs[20]. To account for *Kepler*'s observation cadence, we averaged our transit model with 11 equal spacings within the 1 minute or 30 minute integration window. We evaluated the photometric noise for each quarter of data to fit transit models, adopting published stellar parameters for Kepler-138[32]. We adopted a two-parameter quadratic model for limb-darkening with fixed coefficients (0.3576, 0.3487) appropriate for *Kepler*'s bandpass and Kepler-138's effective temperature ($T_{eff}$), log($g$) and metallicity [Fe/H][33].

The light curve model parameters consist of the mean stellar density[34] ($\rho_{star}$), a photometric zero point for each light curve segment, and for each planet the orbital period, time of transit, planet-to-star radius ratio, impact parameter, and eccentricity parameterized as $e\cos\omega$ and $e\sin\omega$. We determined posterior distributions of our model parameters using MCMC techniques[20]. Our best-fit transit models, shown in ED Figure 1, resulted in consistent estimates for $\rho_{star}$ from each planet.

We determined the mass and radius of Kepler-138 by fitting the spectroscopic parameters ($T_{eff}$, [Fe/H])[32] and our light curve contraints of $\rho_{star}$ to Dartmouth Stellar Evolution models[35], assuming a Gaussian probability density for each parameter[32]. For the Dartmouth models, we varied initial conditions of Mass, Age, and [Fe/H] and interpolated over a grid to evaluate $T_{eff}$, $\rho_{star}$, and [Fe/H] for any set of initial conditions. We computed posteriors using MCMC to obtain stellar model-dependent posteriors on $M_{star}$ and $R_{star}$. Table 1 lists our adopted stellar parameters. We tested the effects of eccentricity priors on our measurement of $\rho_{star}$, adopting a uniform prior in eccentricity as our nominal results. We compare results with eccentricity fixed at zero in ED Figure 2. Although a uniform prior on eccentricity results in a slightly wider range of inferred radii for the star, both of these models are consistent with the spectroscopic study of Kepler-138[32].

Our solution for the impact parameter of the middle planet is 0.3 +/- 0.2, significantly lower than the previous estimate of 0.92 +/- 0.02[21]. The apparent U-shaped transit for planet 'c' (ED Figure 1) is consistent with a low impact parameter. Our measured impact parameter for planet 'd' is 0.810 +/- 0.057, consistent with the previous measurement[21]. Our revised impact parameters imply $\rho_{star}$ = 9.0 +/- 1.9 g cm$^{-3}$, agrees with our transit models for each planet, and with the spectroscopic study of Kepler-138[32].

We find that Kepler-138 has a smaller mass and radius than previous estimates based on the absolute K-band magnitude ($M_K$) of the star from high-resolution Keck spectra[36]. These relied on mass-luminosity relations[37] and a mass-radius relation from interferometry[38]. However, the calibration stars used to correlate the $M_K$ to spectral index excluded cool stars that are active. In the case of Kepler-138, the photometric time-series exhibits large (1%) variations due to starspots. These increase the risk of systematic errors in the measurement of stellar luminosity, and therefore the stellar properties derived from the mass-luminosity relation.

The time-scale for star-spot modulation, ~20 days, was much longer than the transit duration, and was likely dominated by two spots. We found no evidence of star-spot crossings, nor did we find any TTV periodicities related to the rotation period of the star. Hence, the stellar activity is unlikely to effect our transit model or transit times.

**Analytical Constraints from Transit Timing Variations**
Orbital period ratios determine how close planets are to mean motion resonance, and over what period TTVs are expected to cycle. The inner pair of planets of Kepler-138 orbit near the 4:3 resonance with an expected TTV cycle of 1570 days, slightly longer than the 1454-day observational baseline of Kepler-138 b's transits. We fitted a sinusoid at this periodicity to the TTVs of the inner planet, and detected a TTV amplitude of 34 +/- 4 minutes. This permits a rough estimate for the mass[13] of the middle planet of 6.8 +/- 0.9 $M_{Earth}$ ($M_{Sun}/M_{star}$), which is close to our final measure of the mass of the middle planet.

The detection of TTVs at planet 'b' imply that the TTVs in 'c' have a component caused by planet 'b'. However, the middle planet's TTVs are the combined effect of perturbations from its two neighbours 'b' and 'd', and the outer pair orbit near a second order mean motion resonance for which there is no known analytical model. Hence, the masses of the inner and outermost planets cannot be estimated by fitting such a simplified sinusoidal model to the TTVs.

**Detailed TTV Modelling**
For each set of initial conditions, we calculate the transit times of all three planets based on Newtonian gravity, using an eighth order Dormand-Prince Runge-Kutta integrator[15,16,39]. We compare the simulated and observed transit times for each planet assuming each observed transit time has an independent Gaussian measurement uncertainty.

We found an excess of outlying transit times to our model, where either instrumental effects or stellar activity led to a few unlikely transit times with underestimated uncertainties. ED Figure 3 shows the distribution of residuals to our best-fit TTV model compared to a Gaussian distribution, revealing these outliers. Of the 257 measured transit times, 5 are outliers where for simulated transit times S, and measurement uncertainties $\sigma_{TT}$, (O-S)/$\sigma_{TT}$ > 3. We removed these outliers, and used the 252 remaining measurements as our nominal dataset for dynamical models. Later, we tested our results for robustness against outliers.

Performing extensive grid searches and Levenberg-Marquardt, we found a single region of high posterior probability including multiple closely related local minima. To characterize the masses and orbital parameters of the three planets, we performed Bayesian parameter estimation using a Differential Evolution Markov Chain Monte Carlo (DEMCMC) algorithm[40,41]. We used Metropolis-Hasting acceptance rules on 45 "walkers" exploring parameter space in parallel, where for each walker, a proposal is a scaled vector between two other walkers chosen at random. Using differential evolution for the proposal steps increases the probability that proposals will be accepted, particularly for target distributions with significant correlations between model parameters.

The walkers were launched near the best-fit model found by Levenberg-Marquardt. We updated the vector scale length factor every 15 generations to keep the acceptance rate near the optimum value of $0.25$[40,41,42] and thinned the data by recording every 20th generation in the Markov Chain. We discarded the first 50,000 generations and continued the Markov Chain for 1,250,000 additional generations.

The mean correlation between parameter values in the same MCMC chain for a given separation in the chain begins near unity for consecutive generations, and declines for greater separations[43]. We used this property to assess how well-mixed our MCMC chains were. The autocorrelation length, defined as the lag between generations required for the correlation to fall below 0.5, varied between the walkers, averaging 9000 generations for the planet-star mass ratio of Kepler-138 b. Hence, the mean length of the DEMCMC chains were ~139 autocorrelation lengths.

For each planet, we adopted a uniform prior in orbital period, $T_0$, e, and ω. For $M_p/M_{star}$, we adopted a uniform prior that allowed negative masses to enable a simple estimate for the significance of a positive mass for Kepler-138 b from the posterior samples[27]. For Kepler-138 b, the mass is greater than zero in 99.84% of posterior samples. As a further test, we considered an alternative more realistic uniform prior where planetary masses are positive definite and limited to the mass of a pure iron planet given their sizes[44,45].

Posteriors for the mass ratios of each planet to the host star, and each eccentricity vector component are shown in ED Figure 4. We list 68.3%, 95.4%, and 99.7% credible intervals for all parameters in ED Table 2. We also include the mass ratios between the planets and relative eccentricity vector components, which are more constrained by the data than absolute masses and eccentricities.

**Orbital Eccentricity**
Joint posteriors for the eccentricity vector components are displayed in ED Figure 5. These show extreme correlations due to a broad class of orbital models satisfying the data, in which there is a precise apsidal alignment of orbits.

We performed long-term integrations on a subset of our posterior planet masses and orbital parameters including a wide range of eccentricities using the HNBODY code[46]. We investigated whether long-term stability could further constrain the eccentricities. A sample of the solutions were integrated for 10 million orbits and all were found to be stable. Integrating one of the best-fit solutions with high eccentricities for b, c and d at 0.23, 0.20 and 0.18 respectively, we confirmed that the orbits were stable for over 1 Gyr; the apsidal lock was maintained with periapses of all three planets closely aligned. Although rare in the Solar System, apsidal alignment has been observed and studied in the Uranian ring system[47,48] and it has been detected in exoplanetary systems, including υ Andromedae[49], GJ 876[50] and possibly 55 Cancri[51].

**Significance of Mass Detection for Kepler-138 b**

We establish the significance of a non-zero mass for Kepler-138 b by computing the Bayes factor, i.e., the ratio of the marginalized posterior probability for a model with the mass of planet Kepler-138 b fixed at zero relative to the marginalized posterior probability for a model where all three planets have a non-zero mass. Intuitively, the Bayes factor quantifies how much the transit timing data has increased our confidence that planet b has a non-zero mass. When performing Bayesian model selection, it is essential to choose proper (i.e., normalized) priors for any parameters not occurring in both models. For the mass of Kepler-138 b we adopt a uniform prior ranging between zero mass and the mass of an iron sphere the size of Kepler-138 b. We tested two models of iron planets as upper limits on our mass priors[44,45]. We compute the Bayes factor using the generalized Savage-Dickey density ratio based on the posterior samples from our nominal model[52]. We find that the three massive planet model is strongly favored, with the posterior probability for the three massive planet model equal to 99.82%[44] (99.80%[45]) for our nominal model (i.e., three massive planets, each with a uniform eccentricity prior) or 99.91%[44] (99.90%[45]) for a model with three massive planets, each with a Rayleigh distribution (sigma = 0.02[53]) for an eccentricity prior. A more restrictive prior for the mass of Kepler-138 b would further increase the posterior probability for the three massive planet model.

The generalized Savage-Dickey density ratio (SDDR) is superior to more commonly used substitutes (e.g., Akaike information criterion, AIC, or Bayesian Information Criterion, BIC), since the SDDR provides a practical means for calculating the Bayes factor, the actual quantity of interest for rigorous Bayesian model comparison. The AIC and BIC use only the likelihood of the two best-fit models and do not account for the width or shape of the posterior probability distributions. ED Figure 4a shows that the marginal posterior for the mass of planet 'b' is asymmetric and non-Gaussian, so the AIC or BIC would be a particularly poor choice for our problem.

Therefore, we have computed the rigorously correct Bayes factor using the SDDR, which provides an efficient way of calculating the Bayes factor when comparing two nested models, meaning that the simpler model is equivalent to the more general model when the additional parameters ($\theta$) take on a particular value ($\theta=\theta_0$). In our case, this occurs when $M_b = 0$. One advantage of the SDDR over computing fully marginalized likelihoods is that the SDDR can be computed from the posterior for the more general model. This is computationally practical when comparing models that differ by one to a few dimensions, since then the posterior at $\theta_0$ can be computed from a posterior sample using a kernel density estimator. We estimate the posterior density using a Gaussian kernel density estimator with bandwidth 0.001 $M_{Earth}(M_{star}/M_{Sun})$, which was found to be optimal when analyzing synthetic posterior samples data. We verified that the results were insensitive to varying the choice of bandwidth by an order of magnitude.

**Sensitivity Analyses**
We performed several tests to assess the robustness of our results to: 1) the choice of prior for eccentricity, 2) the treatment of transit time outliers, 3) the assumption of coplanarity, and 4) our algorithm. For these sensitivity analyses, we adopt our nominal prior for planet masses to allow for negative planet masses. While any such models are clearly unphysical, allowing for such model offers an efficient and intuitive means for evaluating whether the lower limit on the planet masses is robust to the above assumptions.

**Choice of Eccentricity Prior**
As noted above, our nominal model has a uniform prior in eccentricity. The joint posterior for the mass ratio of Kepler-138 b to the host star and its orbital eccentricity is shown in ED Figure 5. We show this plot with two alternative priors in eccentricity; a Rayleigh distribution with a scale length $0.1^{54}$, and a more constrained one, consistent with *Kepler*'s multi-planet systems, with a scale length of $0.02^{53}$. Since the data constrain the eccentricity so weakly, the choice of priors strongly affects the posteriors of eccentricity. However, the planet-star mass ratios were much more weakly affected by the choice of eccentricity prior, as shown in ED Figure 7a.

Because the apsidally-locked solutions are long-term stable, we adopt the uniform prior as our nominal solution, and note that a more constraining prior on eccentricity results in a marginally wider posterior for mass ratio of Kepler-138 b to the star.

Although, as noted above, the orbital eccentricities in the system are only weakly constrained, the relative eccentricities and mass ratios between the planets are tightly constrained by the TTVs, as shown in ED Figure 6 and ED Table 2.

**Transit Time Observation Outliers**
To assess the effect of residual outlying transit times, we repeated the analysis with two other sets of measured transit times, differing only in that we removed transit time residual outliers beyond *i)* 4σ (leaving 254 transit times) or *ii)* 2.5σ (leaving 244 transit times), as opposed to our nominal dataset which excluded 3σ outliers. ED Figure 3 shows the relative frequency of residuals compared to a Gaussian. The majority of outliers are between 2.5 and 3σ. We excluded outliers beyond 4σ from all models.

The posterior for the mass of Kepler-138 b for each of these three datasets is displayed for comparison in ED Figure 7b. Overall, the outliers have a modest effect on the measured mass ratio for Kepler-138 b to the host star. Since most of the outliers are in the transit times of Kepler-138 b, these have a small effect on our mass measurement of planet 'b'. Furthermore, the mass of planet 'c' is constrained by its effect on the transit times of both planets 'b' and 'd'. Nevertheless, we note that the inclusion of more outliers increases the skewness of the posterior for planetary mass and causes the mode of the distribution to shift to a slightly lower mass.

**Assumption of Co-planarity**
The known low inclination dispersion amongst *Kepler*'s multiplanet systems makes coplanarity a reasonable assumption for TTV modeling[55,56,13]. Furthermore, geometric considerations make multi-planet transiting less likely to be observed for systems with large mutual inclinations. Nevertheless, we tested the effect of mutual inclinations on our solutions. We performed two additional sets of simulations: *i)* with the longitude of the ascending node of 'b' as a free parameter, leaving the other two planets as coplanar, and *ii)* with a free ascending node for 'd'. In each case, we adopted a uniform prior for the ascending node. In ED Figure 7c, we compare the posteriors for the mass ratio of Kepler-138 b to the host star with mutual inclinations to our nominal result. The nominal result gives a consistent, but slightly wider posterior than with free ascending nodes.

**Tests with Synthetic Transit Times**
We evaluated our method by generating synthetic datasets of transit times with known planetary masses and orbital parameters to test how well the input parameters were recovered. The results are shown in ED Figure 8. We generated synthetic transit times based on the median values of each parameter from the marginal posteriors of our nominal model and added Gaussian noise to each observation with a standard deviation equal to the measured timing uncertainties. Our analysis of the simulated data results in a posterior for the mass of Kepler-138 b consistent with the true values. While the resulting posterior for the mass of Kepler-138 b shows a slightly higher mode and is less skewed than the posterior for the nominal model, the differences are comparable to the minor effects of transit timing outliers or choice of eccentricity prior. This result validates our both our transit timing method and our TTV analysis.

Additionally, we generated eight independent synthetic datasets with zero mass for Kepler-138 b and other parameters based on our nominal model (ED Figure 8). In all eight cases, the posterior probability for planet b's mass was insignificant, with only 16% to 86% of posterior samples having a mass greater than zero, consistent with expectations for non-detections. This is in sharp contrast to our analysis of the actual data, which results in 99.84% of the posterior yielding a positive mass for Kepler-138 b.

**Planetary Characteristics**
Our adopted credible intervals in planetary mass and density were calculated by repeatedly multiplying samples from the posteriors of planet-star mass ratios, and $M_{star}$. Uncertainties in planetary radii were calculated with the fractional uncertainty in the stellar radius and the uncertainty in planet-star radius ratio added in quadrature. Time-averaged incident flux for each planet compared to the Earth was calculated in the low eccentricity limit, although we note that if the orbits were highly eccentric, the fluxes would be marginally higher.

31. Mandel, K. & Agol, E. Analytic Light Curves for Planetary Transit Searches. *Astrophys. J.* **580**, 171-175 (2002).
32. Muirhead, P.S. *et al.* Characterizing the Cool Kepler Objects of Interest. New Effective Temperatures, Metallicities, Masses and Radii of Low-mass Kepler Planet-candidate Host Stars. *Astrophys. J.* **750**, 37-42 (2012).

**Extended Data Table 1: Transit times of Kepler-138.** All times are expressed in days since Barycentric Julian Day: 2,454,900. Estimated uncertainties give 68% confidence limits. Outliers beyond 3σ in the residuals of dynamical fits are marked with an asterisk.

**Extended Data Table 2: Confidence intervals from distributions found with DEMCMC TTV analysis.** We include the parameters of our dynamical fits, as well as the mass ratios and relative eccentricity vector components between the planets, which have tighter constraints than the absolute masses or eccentricity vector components.

Extended Data Figure 1: Folded light curves with corrections for observed transit timing variations for Kepler-138. The scattered points are photometric relative fluxes and the curves are analytical models of the transit shape described in the text. Panels 'a', 'b' and 'c' correspond to Kepler-138 b, c and d respectively.

Extended Data Figure 2: Stellar mass and radius models using constraints on the stellar mean density inferred from the light curve. In cyan are models that adopted a uniform prior in eccentricity, and in magenta constraints found with orbital eccentricities fixed at zero. Grey points with error bars mark stellar parameters found in the literature whilst the black error bars mark our adopted solution for stellar mass and radius.

Extended Data Figure 3: The distribution of residual normalized deviations from our best fit dynamical model to the raw transit times. The histogram marks deviations (O-S)/$\sigma_{TT}$ where O is the observed transit time, S is the simulated transit time and $\sigma_{TT}$ is the measurement uncertainty. The curve marks a Gaussian distribution.

Extended Data Figure 4: Posterior distributions for TTV model parameters. The planet-star mass ratios ($M_p/M_{star}$) for each planet of Kepler-138 b, c, and d are shown in panels a, b, and c, $e\cos\omega$ (in panels d,e and f), and $e\sin\omega$ (in panels g,h, and i) respectively for our nominal model. The relative frequency for each histogram is scaled to the mode.

Extended Data Figure 5: Joint posteriors of model parameters and the effects of eccentricity priors. The dark (light) grey marks the 68.3% (95.4%) credible intervals for each joint posterior. Panels 'a', 'b' and 'c' plot $M_p/M_{star}$ and eccentricity vector components for the inner

and middle planets, whilst panels 'd', 'e', and 'f' plot the same for the middle and outer planets. Panels g, h and i compare $M_p/M_{star}$ for Kepler-138 b only and its orbital eccentricity, for three eccentricity priors, a uniform prior on eccentricity (g), and models with a Rayleigh distribution of scale factor 0.1 (h), and 0.02 (i).

Extended Data Figure 6: Posterior distributions for mass ratios and relative eccentricities between planets. The mass ratio of the inner and middle planets is shown in panel 'a', and relative eccentricity vector components (the difference in $e\cos\omega$ ($e\sin\omega$) in the inner pair in panel b (c)). The mass ration of the middle and outer planets are plotted in panels 'd', relative eccentricity vector components (the difference in $e\cos\omega$ ($e\sin\omega$) in the outer pair in panel e (f)).

Extended Data Figure 7: Sensitivity tests for the effects of eccentricity prior, outlying transit times and free inclinations on the mass of Kepler-138 b relative to the host star. Panel 'a' compares a uniform prior (black curve, our nominal posterior for all comparisons) and a Rayleigh Distribution with scale factors 0.1 (navy) and 0.02 (cyan). Panel 'b' compares posteriors with 3σ outliers excluded (black), with two alternatives; 4σ outliers (blue) and 2.5σ outliers removed (light green). Panel 'c' compares our nominal model with one with a free ascending node for the inner (purple) or outer (red) planet.

Extended Data Figure 8: Validation of our method with synthetic datasets. The green curve marks the posterior for a synthetic dataset generated with the same parameters as the medians of our nominal posteriors (in Table 1). The agreement between the green and black curves validates our method and our claim for a positive mass detection for Kepler-138 b. The magenta and purple shades are posteriors for models using data generated with zero mass for Kepler-138 b. These zero-mass synthetic models all reproduced non-detections.

– 1 –

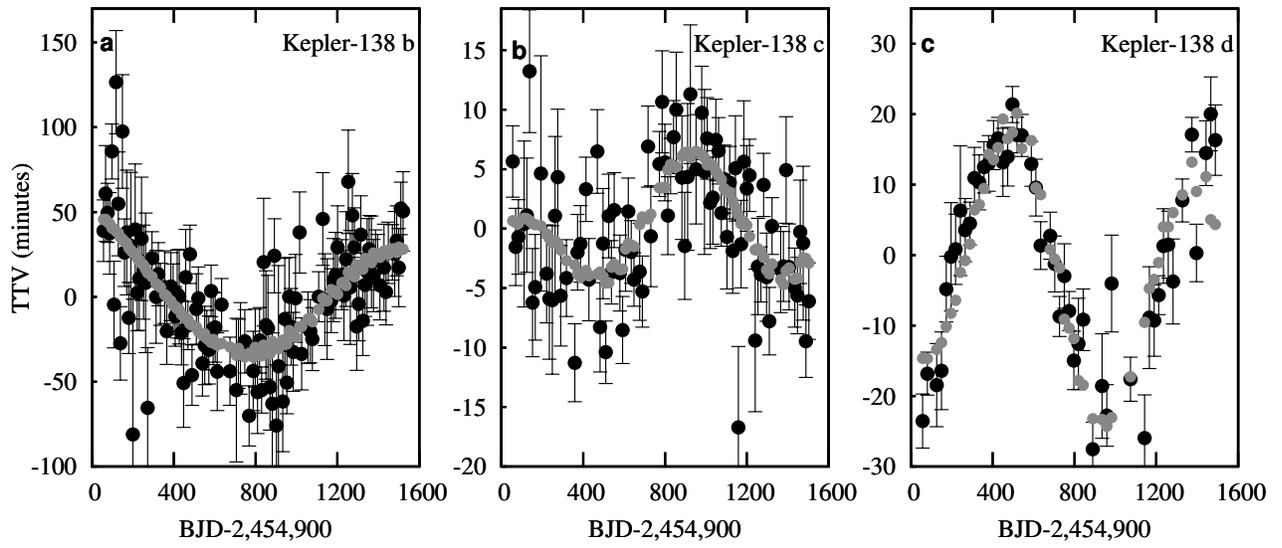

Fig. 1.— Main paper, Figure 1

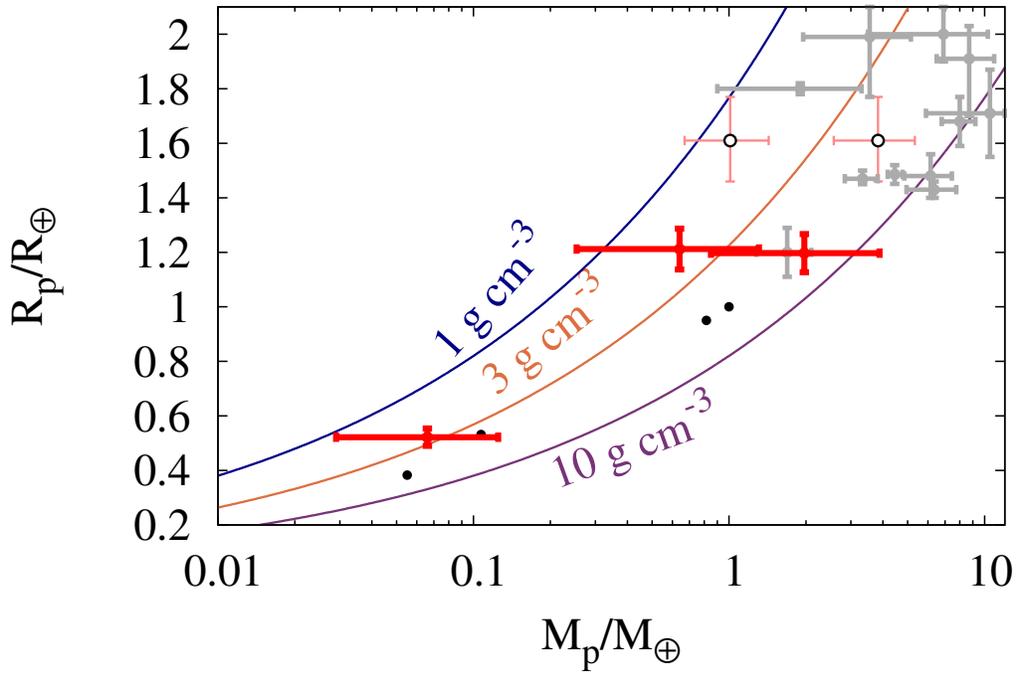

Fig. 2.— Main paper, Figure 2

| Kepler-138 b | | | | Kepler-138 d |
|---|---|---|---|---|
| 56.23045±0.01261 | 633.73803±0.01076 | 1324.71431±0.01427 | 648.31255±0.00155 | 57.81311±0.00267 |
| 66.55886±0.01962 | 674.96352±0.01399 | 1335.04216±0.01297 | 675.87519±0.00283 | 80.90677±0.00212 |
| 76.86414±0.01211 | 705.89533±0.02944 | 1345.36064±0.02184 | 689.65512±0.00208 | 127.08360±0.00410 |
| 87.16890±0.01670 | 716.22338±0.01418 | 1355.68308±0.01380 | 717.22575±0.00237 | 150.17399±0.00382 |
| 97.51565±0.01126 | 726.53871±0.01875 | 1365.99107±0.01877 | 731.00161±0.00295 | 173.27102±0.00324 |
| 107.76610±0.01772 | 747.16800±0.01297 | 1376.29867±0.01476 | 772.34908±0.00190 | 196.36319±0.00534 |
| 118.17035±0.02111 | 757.44480±0.00921* | 1386.61821±0.01230 | 786.13380±0.00298 | 219.45290±0.00346 |
| 128.43371±0.02127 | 767.76381±0.01237 | 1396.93177±0.01012 | 799.91135±0.00223 | 242.54570±0.00639 |
| 138.68975±0.01501 | 788.40859±0.01624 | 1407.23423±0.01422 | 813.68933±0.00227 | 265.63275±0.00311 |
| 149.08968±0.02329 | 798.73107±0.01595 | 1427.86777±0.01784 | 827.47331±0.00161 | 288.72239±0.00210 |
| 159.35323±0.01381 | 809.02624±0.01691 | 1438.17099±0.01347 | 841.25609±0.00213 | 311.81585±0.00301 |
| 169.67471±0.02554 | 819.36067±0.01339 | 1448.50021±0.01257 | 855.03878±0.00336 | 334.90440±0.00272 |
| 179.95288±0.01453 | 829.65344±0.00945 | 1469.12674±0.01137 | 882.59698±0.00361 | 357.99487±0.00252 |
| 190.29561±0.03001 | 840.01903±0.02614 | 1479.43924±0.01741 | 896.37408±0.00312 | 381.08423±0.00274 |
| 200.53153±0.05541 | 850.30643±0.02219 | 1489.75786±0.01432 | 910.15919±0.00428 | 404.17500±0.00242 |
| 210.92867±0.02677 | 860.61829±0.02166 | 1500.06004±0.01604 | 923.94511±0.00405 | 427.26468±0.00206 |
| 221.21582±0.01553 | 870.90735±0.01757 | 1510.39743±0.01074 | 951.50290±0.00319 | 450.35127±0.00336 |
| 231.53491±0.02127 | 881.21371±0.02906 | 1520.70942±0.01625* | 979.06837±0.00273 | 473.44082±0.00289 |
| 241.86431±0.02522 | 891.58748±0.01517 | Kepler-138 c | 992.84600±0.00482 | 496.53494±0.00178 |
| 252.16382±0.01926 | 901.83111±0.02494 | 55.73273±0.00208 | 1006.62904±0.00238 | 519.62090±0.00176 |
| 262.47265±0.02855 | 912.16861±0.04122 | 69.50883±0.00225 | 1020.40638±0.00267 | 542.70985±0.00208 |
| 272.73469±0.02464 | 922.61657±0.01640* | 83.29050±0.00217 | 1034.18777±0.00273 | 588.88500±0.00221 |
| 293.42230±0.01085 | 932.78055±0.02060 | 110.85343±0.00484 | 1047.97221±0.00238 | 611.97165±0.00281 |
| 314.03280±0.01904 | 943.12746±0.02492 | 124.63503±0.00333 | 1061.75267±0.00194 | 635.05492±0.00235 |
| 324.35524±0.01297 | 953.41465±0.01273 | 138.42452±0.00358 | 1075.53011±0.00193 | 681.23384±0.00232 |
| 334.66018±0.01269 | 963.76298±0.01606 | 152.19210±0.00315 | 1089.31318±0.00199 | 704.32004±0.00276 |
| 345.02249±0.01162* | 974.06380±0.01566 | 165.97408±0.00308 | 1103.09090±0.00304 | 727.40385±0.00198 |
| 355.28950±0.01491 | 984.36675±0.01611 | 193.54290±0.00687 | 1116.87516±0.00212 | 750.49681±0.00318 |
| 365.58461±0.01360 | 994.70183±0.01800 | 221.09921±0.00286 | 1130.65224±0.00246 | 773.58233±0.00193 |
| 386.22924±0.01392 | 1015.35516±0.01660 | 234.87885±0.00353 | 1144.43817±0.00307 | 796.66645±0.00288 |
| 396.53721±0.01443 | 1025.61860±0.01477 | 248.65986±0.00433 | 1158.20412±0.00472 | 819.75707±0.00448 |
| 406.84333±0.01218 | 1056.56881±0.01466 | 262.44586±0.00462 | 1171.99592±0.00255 | 842.84842±0.00305 |
| 417.16572±0.01807 | 1066.88059±0.01294 | 276.22920±0.00397 | 1185.78181±0.00356 | 889.01363±0.00268 |
| 427.47777±0.01928 | 1077.19053±0.01288 | 290.00336±0.00294 | 1199.56134±0.00251 | 935.19781±0.00515 |
| 437.77623±0.01316 | 1108.14741±0.01024 | 317.56657±0.00224 | 1213.34319±0.00211 | 958.28389±0.00303 |
| 448.06880±0.01816 | 1128.80557±0.01898 | 358.90489±0.00228 | 1240.89572±0.00416 | 981.38587±0.00477 |
| 458.42534±0.01954 | 1139.08576±0.01882 | 372.69242±0.00221 | 1254.68113±0.00281 | 1073.73237±0.00217 |
| 468.71871±0.00985 | 1149.39509±0.02059 | 386.47399±0.00224 | 1268.46175±0.00161 | 1142.99351±0.00424 |
| 479.06107±0.01195 | 1170.02484±0.01307 | 414.03936±0.00188 | 1282.24807±0.00185 | 1166.09435±0.00502 |
| 489.32477±0.01240 | 1180.34646±0.01704 | 427.81517±0.00240 | 1296.02377±0.00257 | 1189.18306±0.00352 |
| 509.97820±0.01444 | 1190.66161±0.01759 | 469.16591±0.00245 | 1309.80228±0.00198 | 1212.27453±0.00191 |
| 520.29576±0.01286 | 1200.98622±0.01705 | 482.93674±0.00261 | 1323.58892±0.00166 | 1235.36836±0.00368 |
| 530.64900±0.01046* | 1221.60140±0.01428 | 496.72271±0.00322 | 1364.92964±0.00252 | 1258.45747±0.00330 |
| 540.89551±0.01345 | 1231.90599±0.01844 | 510.49745±0.00182 | 1378.71091±0.00204 | 1281.54282±0.00419 |
| 551.21595±0.01603 | 1242.23411±0.01937 | 524.28649±0.00244 | 1392.49762±0.00311 | 1327.72877±0.00212 |
| 561.53131±0.01220 | 1252.57897±0.02116 | 538.06429±0.00185 | 1406.27306±0.00251 | 1373.91321±0.00170 |
| 571.84046±0.01392 | 1262.84891±0.02335 | 551.84901±0.00219 | 1433.83402±0.00231 | 1396.99052±0.00283 |
| 582.17781±0.01054 | 1273.19155±0.01717 | 579.40750±0.00221 | 1447.61464±0.00223 | 1443.17833±0.00318 |
| 592.47561±0.01480 | 1283.49165±0.01618 | 593.18526±0.00196 | 1461.39945±0.00302 | 1466.27118±0.00365 |
| 602.78923±0.01425 | 1293.77245±0.01799 | 606.97103±0.00241 | 1475.17987±0.00450 | 1489.35757±0.00347 |
| 613.08427±0.01595 | 1304.09485±0.01510 | 620.75436±0.00195 | 1488.95524±0.00212 | 1512.45493±0.00353* |
| | 1314.43639±0.01590 | 634.53308±0.00183 | 1502.73866±0.00223 | |

| Param. \ Conf. Int. | 68.3% | 95.4% | 99.7% |
|---|---|---|---|
| $P$ (days) b | $10.3126^{+0.0004}_{-0.0006}$ | $^{+0.0006}_{-0.0010}$ | $^{+0.0007}_{-0.0014}$ |
| c | $13.7813^{+0.0001}_{-0.0001}$ | $^{+0.0002}_{-0.0002}$ | $^{+0.0002}_{-0.0004}$ |
| d | $23.0881^{+0.0009}_{-0.0008}$ | $^{+0.0016}_{-0.0012}$ | $^{+0.0022}_{-0.0013}$ |
| $T_0$ b | $788.4142^{+0.0027}_{-0.0027}$ | $^{+0.0055}_{-0.0054}$ | $^{+0.0084}_{-0.0081}$ |
| c | $786.1289^{+0.0005}_{-0.0005}$ | $^{+0.0010}_{-0.0010}$ | $^{+0.0015}_{-0.0015}$ |
| d | $796.6689^{+0.0013}_{-0.0013}$ | $^{+0.0025}_{-0.0023}$ | $^{+0.0035}_{-0.0026}$ |
| $e\cos\omega$ b | $-0.011^{+0.096}_{-0.140}$ | $^{+0.273}_{-0.300}$ | $^{+0.383}_{-0.370}$ |
| c | $-0.015^{+0.086}_{-0.126}$ | $^{+0.249}_{-0.289}$ | $^{+0.386}_{-0.390}$ |
| d | $-0.037^{+0.060}_{-0.092}$ | $^{+0.170}_{-0.210}$ | $^{+0.255}_{-0.283}$ |
| $e\sin\omega$ b | $-0.024^{+0.075}_{-0.135}$ | $^{+0.238}_{-0.316}$ | $^{+0.335}_{-0.424}$ |
| c | $-0.020^{+0.064}_{-0.117}$ | $^{+0.205}_{-0.276}$ | $^{+0.293}_{-0.369}$ |
| d | $-0.057^{+0.053}_{-0.095}$ | $^{+0.148}_{-0.226}$ | $^{+0.239}_{-0.292}$ |
| $\frac{m_p}{m_\oplus}\frac{M_\odot}{M_\star}$ b | $0.129^{+0.121}_{-0.078}$ | $^{+0.252}_{-0.108}$ | $^{+0.363}_{-0.131}$ |
| c | $3.846^{+3.771}_{-2.304}$ | $^{+7.431}_{-3.029}$ | $^{+10.003}_{-3.408}$ |
| d | $1.282^{+1.362}_{-0.783}$ | $^{+2.759}_{-1.027}$ | $^{+3.667}_{-1.141}$ |
| $M_b/M_c$ | $0.0353^{+0.0136}_{-0.0125}$ | $^{+0.0284}_{-0.0244}$ | $^{+0.0447}_{-0.0359}$ |
| $e_c\cos\omega_c - e_b\cos\omega_b$ | $0.0038^{+0.0167}_{-0.0139}$ | $^{+0.0389}_{-0.0487}$ | $^{+0.0530}_{-0.0792}$ |
| $e_c\sin\omega_c - e_b\sin\omega_b$ | $-0.0033^{+0.0202}_{-0.0132}$ | $^{+0.0469}_{-0.0379}$ | $^{+0.0625}_{-0.0608}$ |
| $\omega_c - \omega_b$ (°) | $-0.58^{+2.20}_{-2.41}$ | $^{+10.66}_{-10.51}$ | $^{+17.86}_{-16.78}$ |
| $M_c/M_d$ | $2.9586^{+0.4434}_{-0.3528}$ | $^{+1.0303}_{-0.6369}$ | $^{+1.7453}_{-0.8774}$ |
| $e_c\cos\omega_c - e_d\cos\omega_d$ | $0.0312^{+0.0391}_{-0.0205}$ | $^{+0.1059}_{-0.0458}$ | $^{+0.1845}_{-0.0664}$ |
| $e_c\sin\omega_c - e_d\sin\omega_d$ | $0.0186^{+0.0322}_{-0.0319}$ | $^{+0.0874}_{-0.0750}$ | $^{+0.1439}_{-0.1069}$ |
| $\omega_c - \omega_d$ (°) | $3.31^{+9.11}_{-15.30}$ | $^{+13.75}_{-20.34}$ | $^{+14.62}_{-21.24}$ |

Fig. 4.— FD Table 2

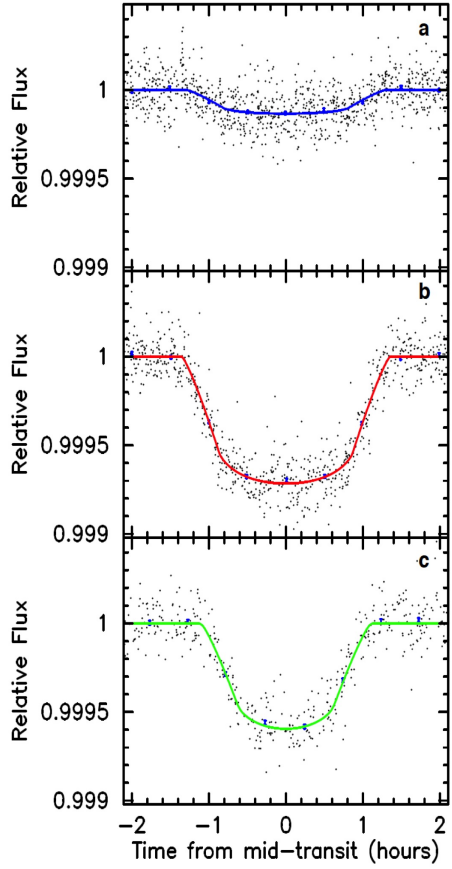

Fig. 5.— ED Figure 1

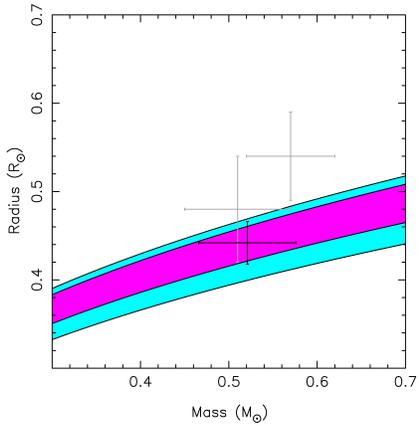

Fig. 6.— ED Figure 2

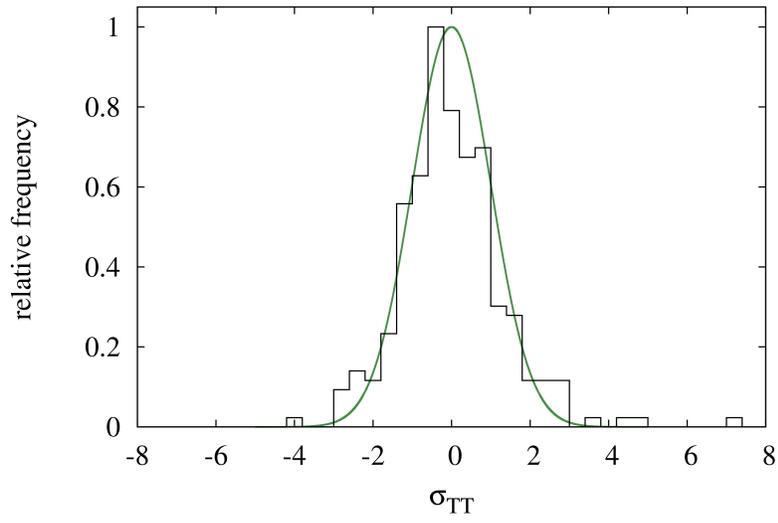

Fig. 7.— ED Figure 3

– 7 –

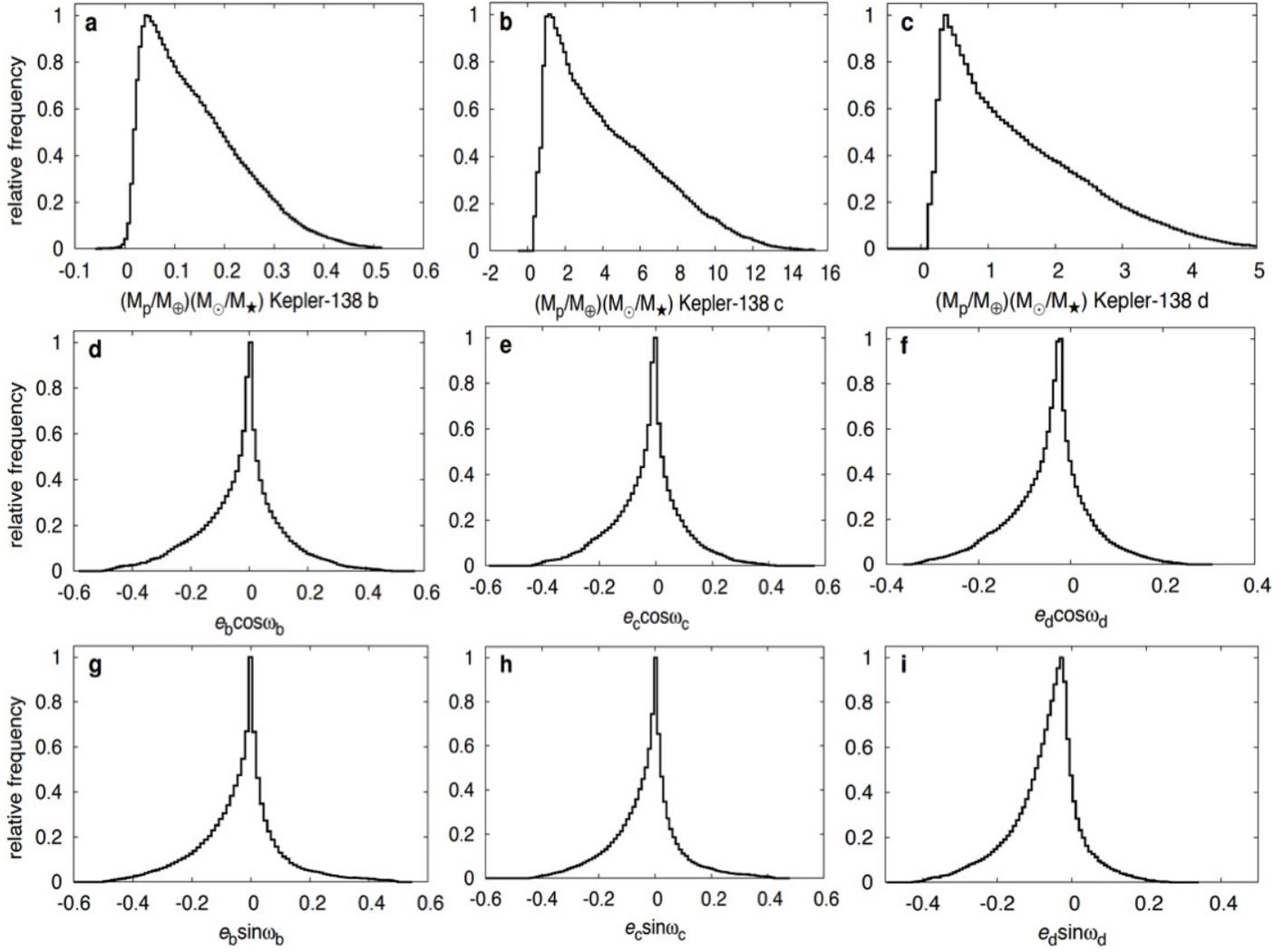

Fig. 8.— ED Figure 4

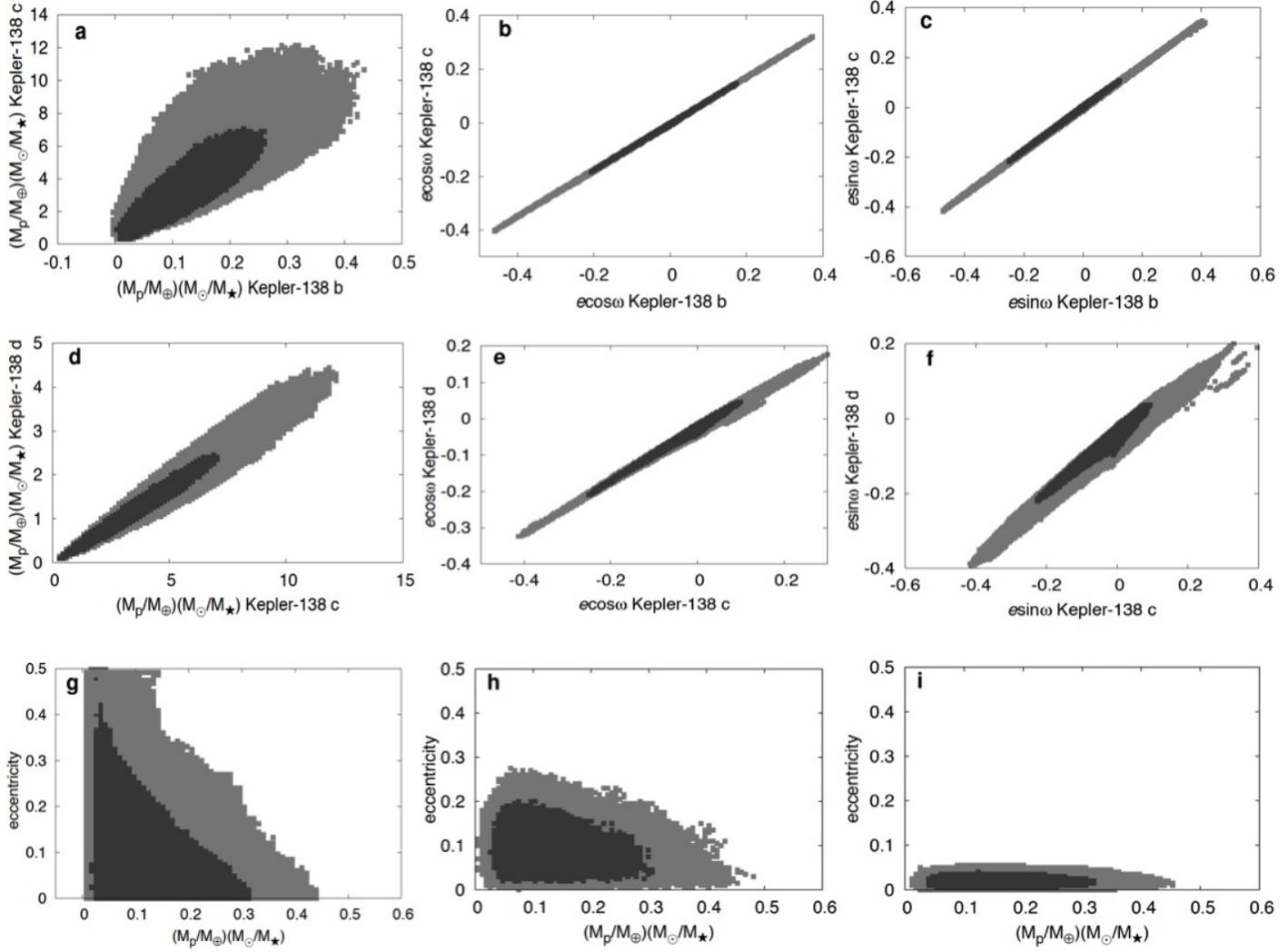

Fig. 9.— ED Figure 5

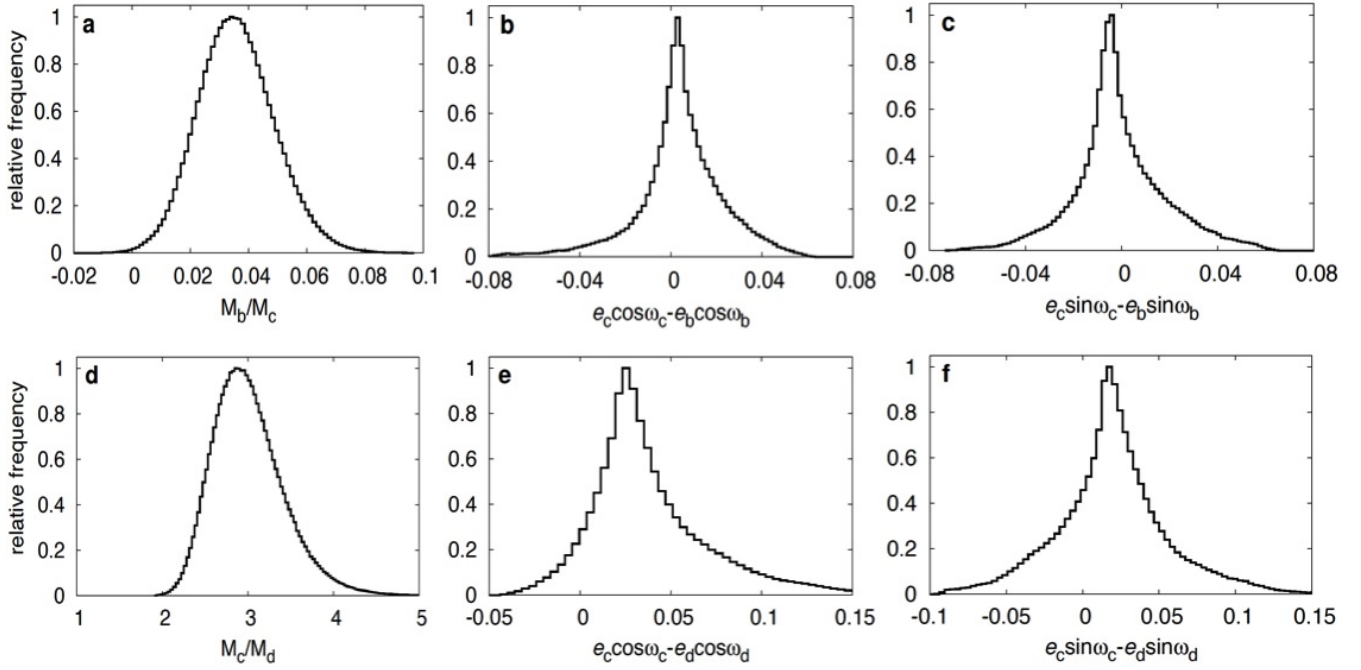

Fig. 10.— ED Figure 6

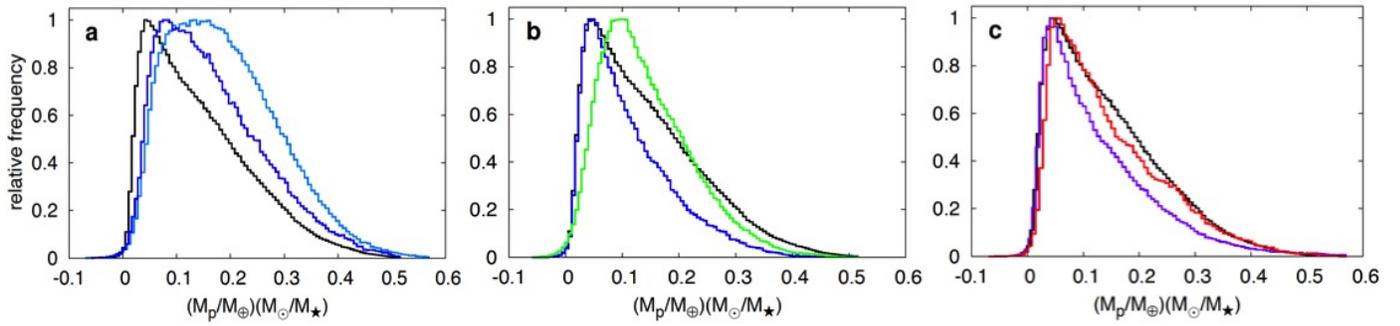

Fig. 11.— ED Figure 7

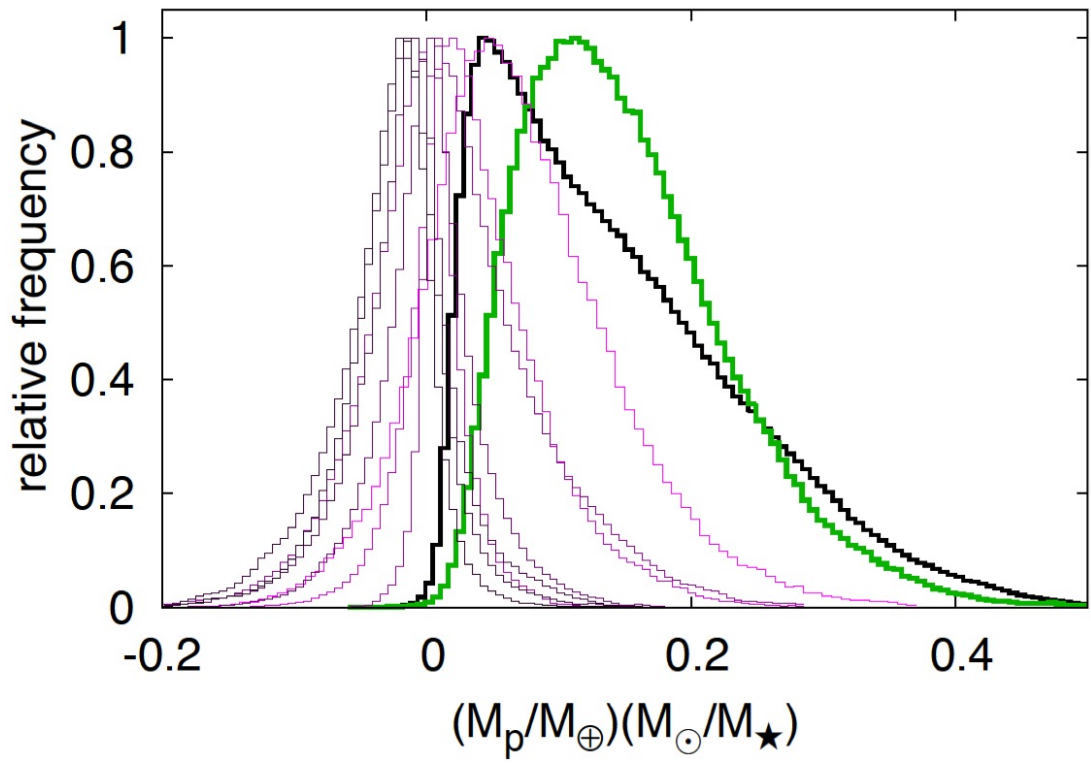

Fig. 12.— ED Figure 8